\documentclass[lettersize,journal]{IEEEtran}
\usepackage{mathrsfs}
\usepackage{amsfonts,dsfont}
\usepackage{amssymb,colortbl}
\usepackage{graphicx}
\usepackage{subfigure}
\usepackage{booktabs,latexsym,verbatim}
\usepackage[font=small,labelfont=bf]{caption}
\usepackage{hyperref}
\usepackage{bm}
\usepackage{cite}
\usepackage{epsfig}
\usepackage{changepage}
\usepackage[fleqn]{amsmath}
\begin{document}

\title{{\color{blue}Scaling properties} of scale-free networks in degree-thresholding renormalization flows}

\author{Dan~Chen, Defu~Cai, and Housheng~Su
\thanks{The work presented was supported by the Science and Technology Project of State Grid Corporation of China, China (5100-202199557A-0-5-ZN). (Corresponding author: Housheng Su.)}
\thanks{Dan Chen and Housheng Su are with the School of Artificial Intelligence and Automation,
Huazhong University of Science and Technology, Wuhan 430074, China,
and also with the Key Laboratory of Image Processing and Intelligent
Control of Education Ministry of China, Huazhong University of Science
and Technology, Wuhan 430074, China.
Email: chend@hust.edu.cn, houshengsu@gmail.com.}
\thanks{Defu Cai is with the State Grid Hubei Electric Power Research Institute, Wuhan, China.
Email: defucai@foxmail.com.}
}

\maketitle

\begin{abstract}
We study the statistical properties of observables of scale-free networks in the degree-thresholding renormalization (DTR) flows. For BA scale-free networks with different sizes, we find that their structural and dynamical observables have similar scaling behavior in the DTR flow. The finite-size scaling analysis confirms this view and reveals a scaling function with a single {\color{blue} scaling} exponent that collectively captures the changes of these observables. {\color{blue}Furthermore, for the scale-free network with a single initial size, we use its DTR snapshots as the original networks in the DTR flows, then perform a similar finite-size scaling analysis. Interestingly,} the initial network and its snapshots share the same scaling exponent as the BA synthetic network. Our findings have important guiding significance for analyzing the {\color{blue}structure and dynamic behavior of large-scale networks}. Such as, in large-scale simulation scenarios with high time complexity, {\color{blue}the DTR snapshot} could serve as a substitute or guide for the initial network and then quickly explore the {\color{blue} scaling} behavior of initial networks.
\end{abstract}


\section{Introduction}\label{1}
\IEEEPARstart{T}{he} network widely exists in both nature and human society and is a common language for describing and modeling complex systems. Its emergence contributes to a better understanding of the structure and functional attributes of the system~\cite{Latora2017,Barabasi2016}. {\color{blue}In this regard, the scale-free network~\cite{Barabasi1999} is undoubtedly an excellent example, and its degree distribution is usually can be described by the power-law distribution $p(k) \sim k^{-\lambda}$ in the mathematical background.} In addition, this characteristic effectively distinguishes scale-free networks from regular networks, {\color{blue}Erd\"os-R\'enyi (ER)} random networks~\cite{Erdos1959}, and {\color{blue}Watts-Strogatz (WS)} small-world networks~\cite{Watts1998}. As a result, scale-free networks have many peculiar properties, such as robustness under random attacks, vulnerability under target attacks~\cite{Albert2000}, and faster spreading speed of virus on scale-free networks~\cite{Pastor2001}.

In the following years, another important property of networks, self-similarity~\cite{Song2005}, has also been discovered by borrowing concepts related to the renormalization group~\cite{Kadanoff2000} in statistical physics. Some major works include: Song {\it et al.}~\cite{Song2005} proposed a box-covering renormalization (BCR) method based on shortest path lengths between network nodes to reduce the size of the network, and they found that the degree distribution of real networks such as WWW remained approximately unchanged during the BCR iteration. Serrano {\it et al.}~\cite{Serrano2008} presented a simple degree-thresholding renormalization (DTR) technique for mining the self-similar properties of real networks. Recently, Garc\'ia-P\'erez {\it et al.}~\cite{Garcia2018} offered a network geometric renormalization (GR) approach in the context of the hidden metric space model~\cite{Serrano2008,Yi2021,Huang2020,Cui2019,Chen2022}, which provides another insight for studying the structural symmetry of networks. {\color{blue}Precisely, the BCR depends on the shortest path length between nodes, and the GR requires embedding the network into a hidden space~\cite{Serafino2021}. Their core idea is to coarse-graining multiple nodes into a supernode to produce smaller-scale replica networks. In contrast, the DTR procedure induces a smaller-scale subgraph by extracting nodes larger than a given degree threshold in the initial network. Technically, the DTR procedure is easier to execute.} In short, the significance of renormalization of the network is to find smaller-replicas to replace the original network, so as to effectively explore the self-similarity of the real network.

{\color{blue}Renormalization is useful for transforming large-scale networks into smaller ones, and the DTR is especially simple and convenient in this regard. Therefore, in the context of DTR, we conduct a series of studies on synthetic and real scale-free networks.} First, we find that the DTR {\color{blue}procedure} can approximately maintain the important structural properties of Barabasi-Albert (BA)~\cite{Barabasi1999} and {\color{blue}Chung-Lu (CL)~\cite{Chung2002,Fasino2021} scale-free networks (see Sec.~\ref{2.2}). Then, our results show that some observables of BA and CL scale-free networks follow the scaling properties in the DTR flow} (see Sec.~\ref{3.1}). The finite-size scaling (FSS) analysis confirms this view and reveals a scaling function to capture the observable's variation (see Sec.~\ref{3.2}). Furthermore, for a CL or real scale-free network with a single initial size, using the DTR procedure to obtain the corresponding smaller-size subnetwork, the results show that the initial network and its subnetworks share the same scaling exponent as the BA network (see Sec.~\ref{3.2}). Finally, we present the conclusion of the paper and the prospect of future work (see Sec.~\ref{4}).

The main contributions of the paper are summarized as follows:

	a) The statistical properties of representative observables of scale-free networks under the DTR procedure are studied and show that these observables obey the scaling law.

	b) {\color{blue}For BA scale-free networks with different initial sizes, the FSS confirmed this universal scaling law and revealed a scaling function with a single exponent ($\alpha = 1$) to capture the behavior of observables in DTR flows. For CL scale-free networks with different initial sizes, our results show that the scaling exponent depends on its degree distribution exponent $\lambda$.}

	c) Finally, for the {\color{blue}CL or real-world scale-free network with a single initial size, our results show that the initial network and its DTR snapshots share the same scaling exponent as the BA synthetic network.}

	d) From the perspective of the application, the scaling exponent obtained here can be used as the foundation for predicting the structure and dynamic characteristics of the large-scale network, which has important guiding significance for reducing the time complexity of the large-scale numerical simulation.

\section{Preliminaries}\label{2}
This section briefly introduces the BA and CL scale-free network model, then reviews the degree-thresholding renormalization (DTR) program. 
The statistical characteristics of their complementary cumulative degree distribution, degree-dependent clustering coefficient, and degree-degree correlations in DTR flows are further studied.

\subsection{Network models}\label{2.1}
{\color{blue}As one of the most classic network model, the BA scale-free network is of great significance for exploring scale-free characteristics and evolutionary mechanisms of networks.} Its generating mechanisms are dominated by two factors: growth and preferential attachment~\cite{Barabasi1999}. In this context, the network eventually develops into a scale-free network with a degree distribution exponent $\lambda \approx 3$. The exact form of the degree distribution of the BA model is $p(k) = [2 m(m+1)]/[k(k+1)(k+2)]$. For large $k$, approximately satisfies $p(k) \sim k^{-3}$, where the parameter $m$ is the number of links increased after adding a new node. The paper uses the Python-based NetworkX library~\cite{Hagberg2008} to generate a BA scale-free network.

\begin{figure}[htbp]
\begin{center}
\includegraphics[width=0.95\linewidth]{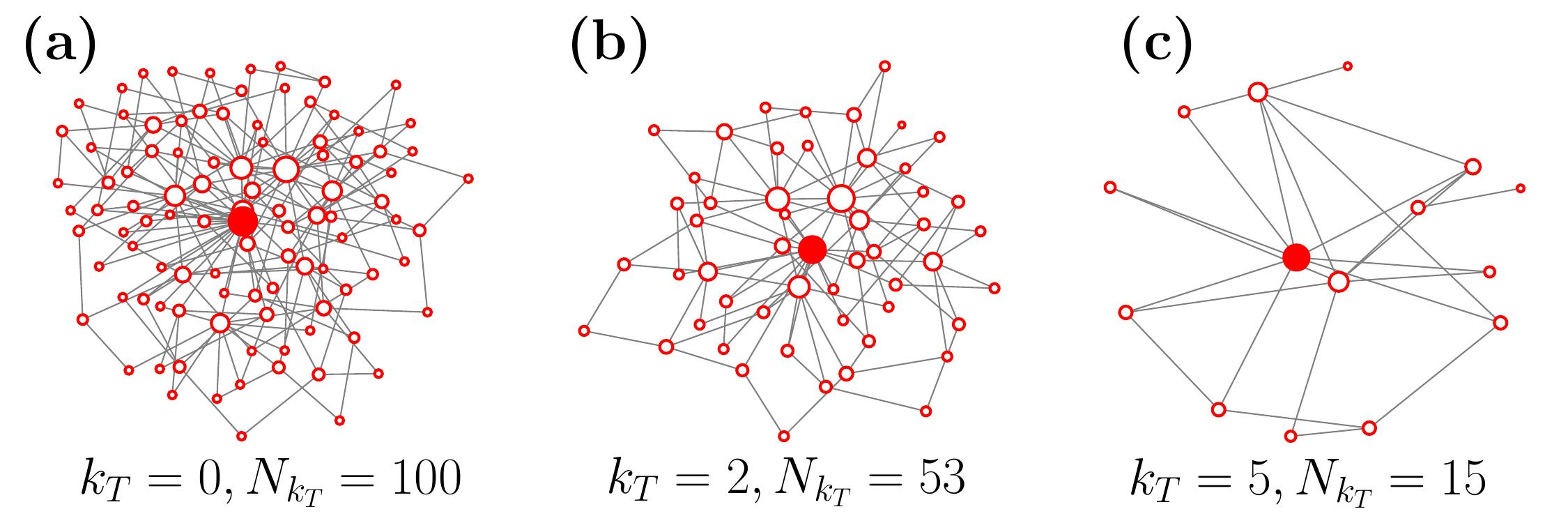}
\caption {Degree-thresholding renormalization diagram. We consider a BA scale-free network with 100 nodes ($k_T = 0$) as the initial network, the parameter $m = 2$. The threshold values $k_T$ = 2, 5 are selected to obtain two network snapshots with progressively smaller-scales.}
\label{Fig:1}
\end{center}
\end{figure}

\subsection{Degree-thresholding renormalization}\label{2.2}
Serrano {\it et al.}~\cite{Serrano2008} proposed a degree-thresholding renormalization (DTR) procedure, and suggested that some basic properties of some real scale-free networks and a class of geometric models are self-similar under this framework, such as the complementary cumulative degree distribution, degree-dependent clustering coefficient, and degree-degree correlations. The specific steps of DTR are as follows: {\color{blue}for an initial network $G_0$ with $N_0$ nodes}, given degree threshold $k_T = 0,1,2, \ldots$, then, nodes with degrees $k > k_T$ are extracted from $G_0$ to obtain the subgraph $G(k_T)$ (i.e., nodes with degrees less than or equal to $k_T$ are deleted), we thereby obtain a series of downscaled subnetworks, and the number of nodes contained in the subnetwork $G(k_T)$ is denoted by $N_{k_T}$. Starting from a BA scale-free network with 100 nodes, we delete those nodes with degrees $k \leqslant k_T = 2,5$ based on the initial network, and obtain two subgraphs (see Fig.~\ref{Fig:1}). The red node in the center is the hub node, which shows that the hub node always exists in the network during this process. 

\begin{figure*}[htbp]
\centering
\includegraphics[width=0.95\linewidth]{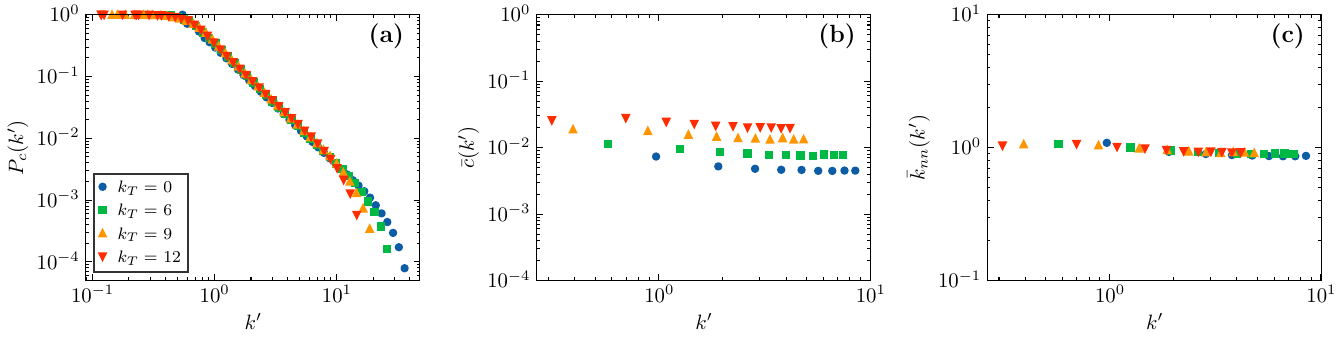}
\caption{{\color{blue}Basic structural characteristics of BA scale-free networks in the DTR flow. (a) Complementary cumulative degree distribution $P_c(k')$, (b) degree-dependent clustering coefficient $\bar{c}(k')$, and (c) normalized average nearest-neighbor degree $\bar{k}_{nn}(k')$. The initial network size is $N_0 = 10^4$, and the average degree satisfies $\langle k \rangle \approx 2 m = 10$. All results are averaged over 100 independent realizations.}}
\label{Fig:2}
\end{figure*}

Figs.~\ref{Fig:2}(a), (b) and (c) show the complementary cumulative degree distribution {\color{blue}$P_c(k)$, degree-dependent clustering coefficient $\bar{c}(k)$, and degree-degree correlations $\bar{k}_{nn}(k)$ of a BA scale-free network and its corresponding subgraphs, respectively. Here $\bar{c}(k)$ is defined as the average of the local clustering coefficients of all nodes with degree $k$. The degree-degree correlation is measured by normalized average nearest-neighbor degree $\bar{k}_{nn}(k) = k_{nn}(k) \langle k \rangle_{k_T} / \langle k^2 \rangle_{k_T}$, and the $k_{nn}(k)$ is defined as the average of the average nearest-neighbor degree of all nodes with degree value $k$, where $\langle k \rangle_{k_T}$ is the average degree of the subgraph $G(k_T)$. Following the practice in Res.~\cite{Serrano2008,Garcia2018}, we show the horizontal axis here with the rescaled degree $k' = k/\langle k \rangle_{k_T}$. For $P_c(k')$ and $\bar{k}_{n n}(k')$, the results of different subgraphs can collapse on almost the same main curve, which indicates that these two characteristics of BA networks are self-similar in the DTR flow. For $\bar{c}(k')$, the curves of different subgraphs do not collapse well. On the contrary, these curves seem to move upward with the increase of $k_T$, which also seems to reflect the essential difference between the BA scale-free network and some self-similar synthetic scale-free networks (such as $\mathbb{S}^1$ model~\cite{Serrano2008}). We only show the result of $m = 5$ here, and a consistent conclusion can be obtained when the parameter $m$ is set to other values (the results can be viewed in the main sample code$\footnote{The main sample code used in this paper can be accessed online in the GitHub public repository: \url{https://github.com/cdzqf/DTR}}$).}

\begin{figure*}[htbp]
\centering
\includegraphics[width=0.95\linewidth]{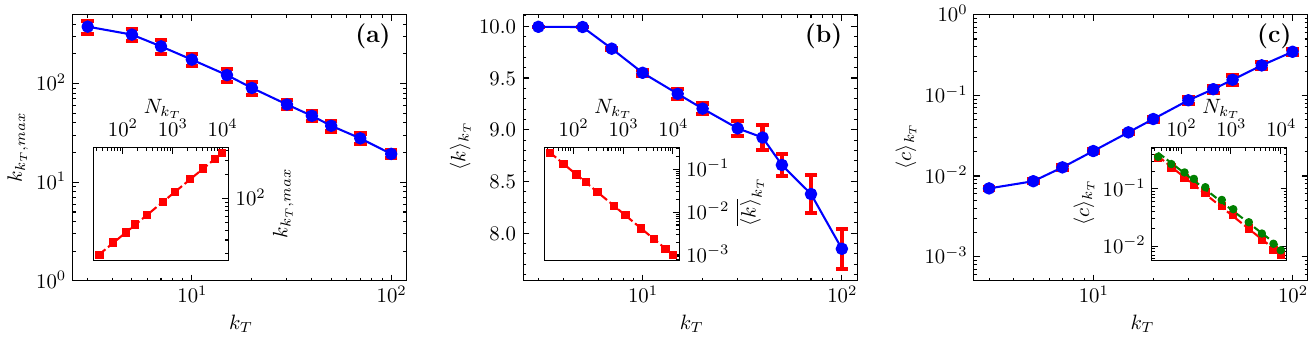}
\caption{Observables of the BA scale-free network as a function of the degree threshold $k_T$,
the inset shows the dependence of each observable on the size of the subgraph $G(k_T)$.
(a) The largest degree $k_{k_T, max}$ of the subnetwork $G(k_T)$ as a function of the degree
threshold $k_T$, the inset shows $k_{k_T, max} \sim N_{k_T}^{\beta}$, where $\beta \approx 0.5228$. 
{\color{blue}Here, the tool provided by Refs.~\cite{Clauset2009,Alstott2014} is used to estimate the degree distribution exponent of the BA scale-free network and obtains $\lambda \approx 2.89$, obviously, $\beta \approx 1/(\lambda - 1)$.} (b) The average degree $\langle k \rangle_{k_T}$ of the subnetwork $G(k_T)$ as a function of the degree threshold $k_T$. The inset shows the dependence of the normalized average degree
$\overline{\langle k \rangle}_{k_T}$ on $N_{k_T}$,
approximately satisfies $\overline{\langle k\rangle}_{k_T} \sim N_{k_T}^{-\gamma}$.
(c) The average clustering coefficient $\langle c \rangle_{k_T}$ of the subnetwork $G(k_T)$
as a function of the degree threshold $k_T$, the inset shows
$\langle c \rangle_{k_T} \sim N_{k_T}^{-\mu}$, the red square is the simulation result, and the green circle is the theoretical result.
The size of the initial network is $N = 10^4$, the parameter $m = 5$, error bars show the standard
deviations for different realizations, and all results are averaged over 100 independent realizations.}
\label{Fig:3}
\end{figure*}

{\color{blue}In addition, compared with the real scale-free network, BA scale-free network is a special case. To this end, we employ a more general scale-free network model, Chung-Lu (CL) scale-free network~\cite{Chung2002,Fasino2021}, as an extension model, its degree distribution satisfies $p(k) \sim k^{-\lambda}$, and the exponent $\lambda$ is used to control the heterogeneity of the network. We employ the recent algorithm presented by Fasino {\it et al.}~\cite{Fasino2021} to create a series of CL networks with different sizes and degree distribution exponents. Specifically, the model needs to define a nonnegative real vector
\begin{equation}
\mathbf{w}=(w_1, \ldots, w_{N_0})^T, w_i=c/(i+i_0)^p,
\end{equation}
where $i=1,2,\ldots,N_0$, $p = 1/(\lambda - 1)$, $c=(1 - p) \langle \kappa \rangle N_0^p$, $i_0 = (c/\kappa_{max})^{1/p} - 1$, $\langle \kappa \rangle$ and $\kappa_{max}$ are the upper limits of the expected values of the average degree and the largest degree, respectively. We obtain a result similar to the BA network, as shown in Fig.~S1 of the Supplemental Material. Our results show that the BA and CL scale-free networks' degree-dependent clustering coefficient seems to have weak self-similarity in the DTR flow. In contrast, the $\mathbb{S}^1$ model shows good self-similarity (see Fig.~S2 in the Supplemental Material).}

\section{Results and Discussion}\label{3}
\subsection{Scaling properties of scale-free networks along the DTR flow}\label{3.1}
In this subsection, we study the statistical properties of observables of BA scale-free networks, {\color{blue}CL scale-free networks}, and real scale-free networks in the DTR flow. Our results show that these observables approximately obey the scaling law in the DTR flow. Specifically, we investigate some basic and important topological characteristics of the network, such as the largest node degree, average degree (edge density), average clustering coefficient, and the $i$th moment of the degree distribution.

Taking the largest degree as an example, in general, the probability that a node's degree value is equal to or higher than $k_{max}$ can be found in the network is $1/N$~\cite{Barabasi2016}. For power-law degree distribution $p(k) \sim k^{-\lambda}$, {\color{blue}we have $\int_{k_{max}}^{\infty} k^{-\lambda} dk \sim 1/N$, and the natural cutoff follows $k_{max} \sim N^{1/(\lambda-1)}$$\footnote{See equation (4.16) in Chapter~4 of the Network Science online reading document: \url{http://networksciencebook.com/chapter/4\#hubs}}$}. Thus, for a network $G_0$ with $N_0$ nodes, the largest degree satisfies $k_{0, max} \sim N_0 ^ {1/ (\lambda - 1)}$. For BA scale-free networks, Fig.~\ref{Fig:3}(a) shows that the largest degree $k_{k_T,max}$ of subgraph $G(k_T)$ is a monotonically decreasing function as the threshold value $k_T$, and the inset shows that $k_{k_T, max} \sim N_{k_T}^{\beta}$. The result implies that the DTR procedure can maintain the scaling characteristics of the largest node degree of the initial BA scale-free network.

\begin{figure*}[htbp]
\centering
\includegraphics[width=0.95\linewidth]{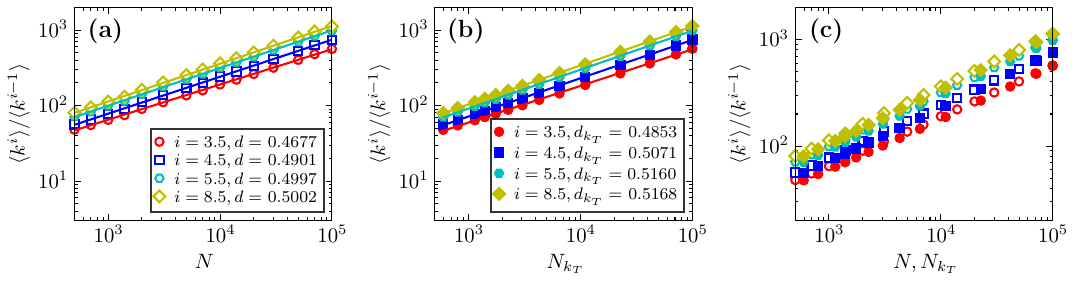}
\caption{Moment ratio test of BA scale-free networks. (a) Generate a series of BA networks with sizes between $[500,10^5]$, examine the dependence of the moment ratio of node degrees on $N$, and the exponent $d$ is the slope obtained by fitting these results. (b) Starting from the BA scale-free network with the initial size of $N = 10^5$, a series of subnetworks are obtained by the DTR procedure, shows the dependence of the moment ratio of the subnetwork on the size $N_{k_T}$, and the exponent $d_{k_T}$ is the slope obtained by fitting. {\color{blue}Here, the degree threshold $k_T$ belongs to the interval $[0, 120]$ to ensure that extracted subnetworks are roughly the same sizes as the series of BA networks generated in Fig.~\ref{Fig:4}(a). (c) Display the results of (a) and (b) in the same coordinate system.} The parameter $m = 5$, and all results are averaged over 100 independent realizations.}
\label{Fig:4}
\end{figure*}

{\color{blue}Fig.~\ref{Fig:3}(b) shows the behavior of the subgraphs' average degree $\langle k \rangle_{k_T}$ in the DTR flow. Different from the results of $\mathbb{S}^1$ geometric models (the average degree increases with the increase of $k_T$)~\cite{Serrano2008}, the BA network's average degree slightly decreases with the increase of $k_T$.} The inset shows the dependence of the normalized average degree $\overline{\langle k \rangle}_{k_T} = \langle k \rangle_{k_T}/[N_{k_T}-1]$ on $N_{k_T}$, which approximately satisfies the power-law behavior, $\overline{\langle k \rangle}_{k_T} \sim N_{k_T}^{-\gamma}$. Fig.~\ref{Fig:3}(c) shows the average clustering coefficient of the subgraph gradually increases along the direction of DTR flow. The results of the inset show that $\langle c \rangle_{k_T} \sim N_{k_T}^{-\mu}$, as shown in the red square, and the dashed line is the fitting result. Klemm {\it et al.}~\cite{Klemm2002} have proved that the average clustering coefficient of the BA network meets $\langle c \rangle \sim (lnN)^2/N$, and the average clustering coefficient of subgraph $G(k_T)$ can be calculated based on this conclusion, as shown in the green circle, {\color{blue}our results show that the green circle and the red square roughly overlap}, which implies that the DTR procedure do not change the statistical law of the average clustering coefficient for the initial BA network.

{\color{blue}We also performed further research on the CL scale-free network and $\mathbb{S}^1$ geometric model, as shown in Figs.~S3 and S4 of the Supplemental Material. For CL models, except for the average degree of subgraphs, the other results are similar to BA. For the CL and $\mathbb{S}^1$ model, the average degree of the subnetwork increases with the increase of $k_T$ [see Fig.~S3(b) and Fig.~S4(b)]. The subnetwork's average degree depends on many factors, among which the structural characteristics of the network and the finite-size effect may be significant factors. In addition, from the results of BA and CL networks, it seems that the DTR cannot well retain the average clustering coefficient of initial networks [see Fig.~\ref{Fig:2}(b), Fig.~\ref{Fig:3}(c), Fig.~S1(b), and Fig.~S3(c)] but can well retain this feature of $\mathbb{S}^1$ models [see Fig.~S2(b) and Fig.~S4(c)].}

\begin{table*}[tb]
\centering
\caption{The basic topological properties of real scale-free networks
and the values of some scaling exponents. From left to right, we report the network name,
category, number of nodes $N$, number of edges $E$, the average degree $\langle k \rangle$, the degree distribution exponent $\lambda$, the largest degree exponent $\beta$, the normalized average degree exponent $\gamma$, and the scaling exponent $\alpha^*$.}
\label{table:1}
\begin{tabular}{lcccccccc}
\hline
Name & Category & $N$ & $E$ & $\langle k \rangle$ & $\lambda$ & $\beta$ & $\gamma$ & $\alpha^*$ \\
\hline
Proteome & Biological & 4100 & 13358 & 6.52 & 2.61 & 0.6871 & 1.3956 & 1.00(1) \\
Internet & Technological & 23748 & 58414 & 4.92 & 2.17 & 0.6752 & 1.3098 & 1.00(4) \\
Caida20071105 & Technological & 26475 & 53381 & 4.03 & 2.17 & 0.7030 & 1.2677 & 1.00(5) \\
Words & Text & 7377 & 44205 & 11.98 & 2.24 & 0.8302 & 1.5289 & 1.0(1) \\
Frenchbookinter & Text & 9424 & 23841 & 5.06 & 2.43 & 0.7948 & 1.3914 & 1.00(5) \\
Japanesebookinter & Text & 3177 & 7998 & 5.03 & 2.28 & 0.8028 & 1.4240 & 1.00(5) \\
Youtube & Affiliation & 124325 & 293342 & 4.72 &  2.49 & 0.7220 & 1.3300 &    1.00(5) \\
Recordlabel & Affiliation & 186758 & 233277 & 2.5 & 2.02 & 0.8798 & 0.9633 & 1.00(5) \\
\hline
\end{tabular}
\end{table*}

\begin{figure*}[htbp]
\centering
\includegraphics[width=0.95\linewidth]{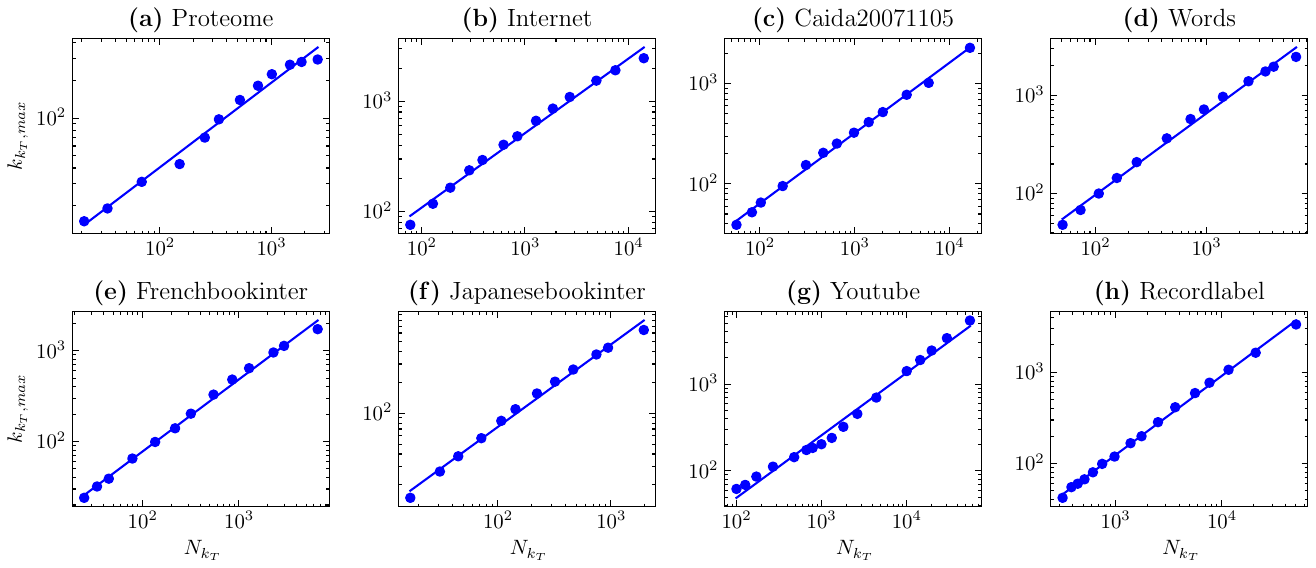}
\caption{For real scale-free networks, shows the dependence of the largest degree $k_{k_T, max}$ of subnetwork $G(k_T)$ on the size $N_{k_T}$, where $k_{k_T, max} \sim N_ {k_T}^{\beta}$, and the value of exponent $\beta$ is shown in Table~\ref{table:1}.}
\label{Fig:5}
\end{figure*}

Recently, Serafino {\it et al.}~\cite{Serafino2021} have shown that finite-size effects usually hide the scale-free properties of real-world networks through finite-size scaling analysis and moment ratio tests. In this context, we employ the moment ratio test to verify the scale invariance of scale-free networks in DTR flows. {\color{blue}The moment of degree distribution is helpful to understand the meaning of scale-free term~\cite{Barabasi2016}. Specifically, for a scale-free network, the $i$th moment of the degree distribution is
\begin{equation}
\left\langle k^i\right\rangle=\int_{k_{\min }}^{\infty} k^i p(k) dk.
\end{equation}
Serafino {\it et al.}~\cite{Serafino2021} have shown that when $i > \lambda$, the ratio of the $i$th moment of the degree distribution to the $(i-1)$th moment is independent of $i$, satisfies $\langle k^{i}\rangle /\langle k^{i-1}\rangle \propto N^d$, where $d > 0$}. As shown in Fig.~\ref{Fig:4}(a), for BA scale-free networks, the moments ratios for different $i$th are parallel lines. Additionally, we consider the BA scale-free network with the initial size of $N = 10^5$ and perform DTR procedure on it to obtain a series of subnetworks. Then, applying the moment ratio experiment to these subnetworks, as shown in Fig.~\ref{Fig:4}(b), we obtain a similar result, namely $\langle k^{i}\rangle /\langle k^{i-1}\rangle \propto N_{k_T}^{d_{k_T}}$, where $d \approx d_{k_T}$. {\color{blue}We display the results of Fig.~\ref{Fig:4}(a) and Fig.~\ref{Fig:4}(b) on the same figure, as shown in Fig.~\ref{Fig:4}(c), the result shows that the DTR procedure can roughly reverse the preferential attachment evolutionary growth process of the BA network, which means that the DTR can approximately return the BA network to an earlier state.}

{\color{blue}According to the results presented in Figs.~\ref{Fig:2}--\ref{Fig:4}, BA networks' degree distribution and degree-degree correlation have scale invariance in the DTR flow. The largest degree, the normalized average degree, and the average clustering coefficient approximately obey the scaling behavior in the DTR flow. These results show that the DTR downscaling process of BA networks can be approximately regarded as the opposite direction of its evolutionary growth process, which makes it possible to predict the structural characteristics of large-scale BA networks based on their small-scale subgraphs.}

In fact, empirical studies show that the degree distribution of a large number of real-world networks approximately satisfies $p(k) \sim k^{-\lambda}$, and in general, the exponent $\lambda \in (2,3)$. Next, we consider eight real-world scale-free networks, which belong to four different categories: Biological, Technological, Text, and Affiliation$\footnote{\url{https://icon.colorado.edu/}.}$. The detailed topological information is shown in Table~\ref{table:1}. After examining the statistical properties of these networks and their subnetworks, respectively, a conclusion similar to Fig.~\ref{Fig:3}(a) is obtained, that is, {\color{blue}the largest degree of the subnetwork satisfies $k_{k_T, max} \sim N_{k_T}^{\beta}$,} as shown in Fig.~\ref{Fig:5}. Furthermore, we also studied the dependence of the normalized average degree of subnetwork $G(k_T)$ on $N_{k_T}$ (detailed results as shown in Fig.~S5). {\color{blue}The results further show that $\overline{\langle k \rangle}_{k_T} \sim N_{k_T}^{-\gamma}$,} where the value of $\gamma$ is shown in Table~\ref{table:1}. {\color{blue}However, our results also show that too large degree thresholds $k_T$ will cause $\overline{\langle k \rangle}_{k_T}$ to deviate from this power-law behavior significantly}, which is obvious on Proteome, Words, Frenchbookinter, Japanebookinter, and YouTube networks. {\color{blue}For these real networks, $\langle k^{i}\rangle /\langle k^{i-1}\rangle \propto N_{k_T}^{d_{k_T}}$ is also approximately true in the DTR flow, as shown in Fig.~S6. Indeed, even if the scale-free network with a sufficient initial size is considered, a large degree threshold will eventually lead to the loss of their original scale invariance (or self-similarity).} For this reason, the finite-size scaling (FSS)~\cite{Stanley1999} analysis in the next section shows that the finite-size effect often masks the underlying scale invariance of many networks.

\begin{figure*}[tb]
\begin{center}
\includegraphics[width=0.95\linewidth]{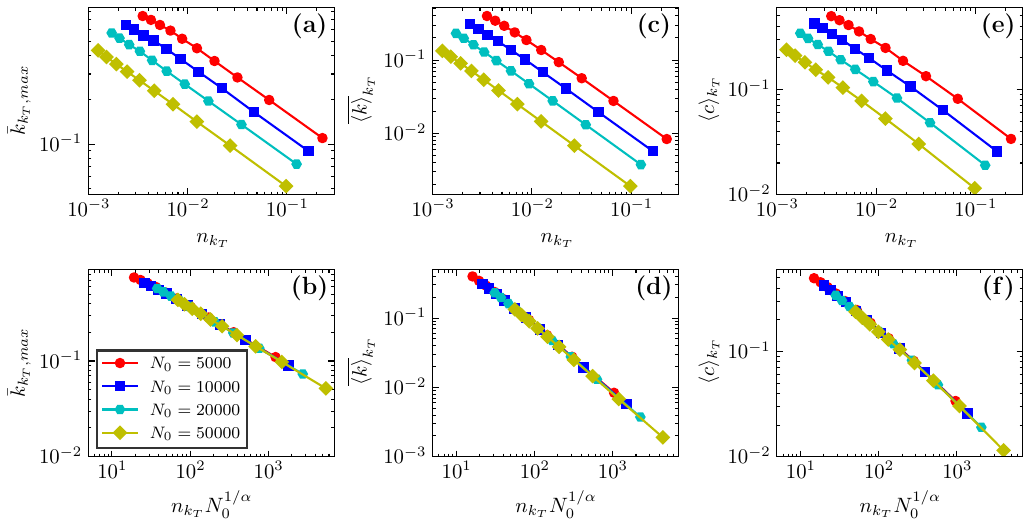}
\caption {Scaling analysis of structural observables for BA scale-free networks with different initial sizes. (a), (c), and (e) show the dependence of each observable on the relative size $n_{k_T}$, respectively. (a) The normalized largest degree $\bar{k}_{k_{T}, \max}$, (c) the normalized average degree $\overline{\langle k\rangle}_{k_{T}}$, and (e) the average clustering coefficient $\langle c\rangle_{k_{T}}$. (b), (d), and (f) show the collapse diagram after rescaling the horizontal axis using the exponent $\alpha \approx 1$, respectively. The parameter $m = 5$, and all results are averaged over 100 independent realizations.}
\label{Fig:6}
\end{center}
\end{figure*}

\subsection{Finite-size scaling analysis}\label{3.2}
For complex systems, some observables deviating from thermodynamic limit behavior are often observed in actual numerical simulations due to the limit of infinite size cannot be reached. Here, we show that the potential scale invariance of network observables is often masked by finite-size effects. We employ the FSS analytical tool to prove this view and reveal a scaling function with a single exponent that jointly captures the scaling behavior of these observables. Using $\mathcal{Z}$ represents a particular observable, we find that under the FSS hypothesis,
\begin{equation}
\mathcal{Z} = f\left(n_{k_T}N_0^{1/\alpha}\right),
\label{Eq:3}
\end{equation}
where $N_0$ is the size of the initial network, $n_{k_T}$ ($n_{k_T} = N_{k_T}/N_0$) is the relative size of the subnetwork $G(k_T)$, and the behavior of the observable $\mathcal{Z}$ is completely determined by the exponent $\alpha$ and the scaling function $f$. {\color{blue}Taking six observables in our recent work~\cite{chen2021} as examples, we next investigate the FSS behavior of these observables in the DTR flow.}

For BA scale-free networks with different initial sizes, Fig.~\ref{Fig:6}(a), (c), and (e) show the dependence of structural observables $\bar{k}_{k_{T}, \max}$ (i.e., ${k}_{k_{T}, \max}/(N_{k_T}-1)$), $\overline{\langle k\rangle}_{k_{T}}$, and $\langle c\rangle_{k_{T}}$ of the subnetwork $G(k_T)$ on the relative size $n_{k_T}$, respectively. {\color{blue}The results show that the observable curves of different-size networks have very similar behavior in the DTR flow. Eq.~\eqref{Eq:3} shows that rescaling the horizontal axis $n_{k_T}$ of these curves to $n_{k_T}N_0^{1/\alpha}$ will cause them to collapse on the same master curve [see Fig.~\ref{Fig:6}(b), (d), and (f)].} Notably, these observables share the same scaling exponent, $\alpha \approx 1$. Here, the optimal exponent $\alpha$ is obtained by measuring the quality $S$ of the collapse plot, where the $S$ is used to measure the mean square distance between the collapse data and the master curve, and its detailed definition is shown in Ref.~\cite{Houdayer2004}. {\color{blue}Precisely, rescaling the horizontal axis $n_{k_T}$ of the original data points to $n_{k_T}N_0^{1/\alpha}$, and the $S$ between multiple rescaled curves is calculated, respectively, where the minimum point of $S$ corresponds to the optimal collapse exponent $\alpha$. Taking the normalized largest degree $\bar{k}_{k_{T}, \max}$ as an example [see Fig.~\ref{Fig:6}(a)], Fig.~S7(a) shows the dependence of the $S$ on exponent $\alpha$. When $\alpha = 0.99$, $S$ achieves the minimum value, which means that different curves have the best collapse [see Fig.~\ref{Fig:6}(b)]. Figs.~S7(b)--S7(d) show the results of Fig.~\ref{Fig:6}(b) at $\alpha=1.0$, $\alpha=0.5$, and $\alpha=1.5$, respectively, and the results show that the overlap quality of Fig.~S7(b) is significantly better than that of Fig.~S7(c) and Fig.~S7(d).}

As a supplement, we also consider some dynamical observables that depend on the underlying structure of the network. Such as the normalized smallest nonzero eigenvalue of the Laplace matrix, $\lambda_{k_T, 2}(L)=\Lambda_{k_T, 2}(L)/ N_{k_T}$, where $\Lambda_{k_T, 2}(L)$ is the smallest nonzero eigenvalue of the Laplace matrix of the subnetwork $G(k_T)$. {\color{blue}Previous studies have shown that, in most cases, maximizing the convergence rate to the network-homogeneous state for undirected networks is equivalent to maximizing $\Lambda_{k_T, 2}(L)$~\cite{Nishikawa2017}}. We further consider the ratio of the largest eigenvalue $\Lambda_{k_T, max}(L)$ of the Laplace matrix to the smallest nonzero eigenvalue $\Lambda_{k_T, 2}(L)$, i.e., $Q_{k_T} = \Lambda_{k_T, max}(L)/\Lambda_{k_T, 2}(L)$~\cite{Nishikawa2003,Donetti2005}. This quantity is related to the stability of the network synchronization process, and it gives the time that the system needs to recover to the stable synchronization state after disturbance~\cite{Barahona2002,Shi2013}. Finally, the largest eigenvalue of the adjacency matrix determines many dynamic behaviors of the network, and its normalized form is $\lambda_{k_T, max}(A)=\Lambda_{k_T, max}(A)/(N_{k_T}-1)$, where $\Lambda_{k_T, max}(A)$ is the largest eigenvalue of the subnetwork's adjacency matrix. Castellano {\it et al.}~\cite{Castellano2017} have shown that the thresholds of two highly correlated dynamic models depend on $\Lambda_{k_T, max}(A)$, one of which is the epidemic spreading, its threshold satisfies $\eta_c = 1/\Lambda_{k_T, max}(A)$, and the synchronization threshold $\zeta_c=\zeta_0/\Lambda_{k_T, max}(A)$~\cite{Restrepo2005} for Kuramoto dynamics coupling parameter associated with $\Lambda_{k_T, max}(A)$. Therefore, these observables (called dynamical observables in this paper) have a crucial influence on the dynamical behavior of the underlying network structure, and investigating their scale invariance in the DTR flow is of great significance for predicting the dynamic behavior of large-scale networks.

\begin{figure}[htbp]
\begin{center}
\includegraphics[width=0.9\linewidth]{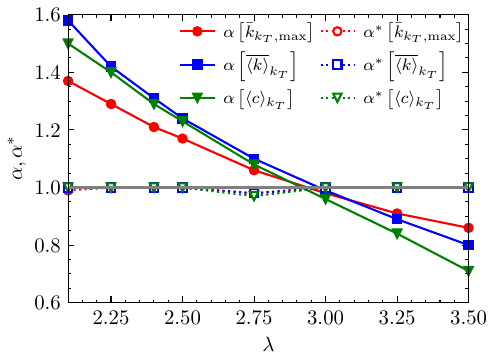}
\caption {{\color{blue}Dependence of scaling exponent $\alpha$ and $\alpha^*$ on heterogeneity exponent $\lambda$. Solid symbols (red solid circle, blue solid square, and green solid triangle) are the results obtained by FSS analysis of CL scale-free networks with different initial sizes (here we consider four initial sizes, $N_0 = 10^4$, $2 \times 10^4$, $5 \times 10^4$, and $10^5$), and the corresponding scaling exponent is $\alpha$. Hollow symbols (red hollow circle, blue hollow square, and green hollow triangle) represent the results of a CL scale-free network ($N_0 = 10^5$) and its subnetworks, and the corresponding scaling exponent is $\alpha^*$. The gray horizontal straight line shows the position of $\alpha = \alpha^* = 1$. $\alpha[\mathcal{Z}]$ and $\alpha^*[\mathcal{Z}]$ represent the scaling exponent of the observable $\mathcal{Z}$ under Eq.~\eqref{Eq:3} and Eq.~\eqref{Eq:4}, respectively, where $\mathcal{Z}$ represents one of $\bar{k}_{k_{T}, \max}$, $\overline{\langle k\rangle}_{k_{T}}$, and $\langle c\rangle_{k_{T}}$. All results are averaged over 100 independent realizations.}}
\label{Fig:7}
\end{center}
\end{figure}

\begin{figure*}[htbp]
\begin{center}
\includegraphics[width=0.95\linewidth]{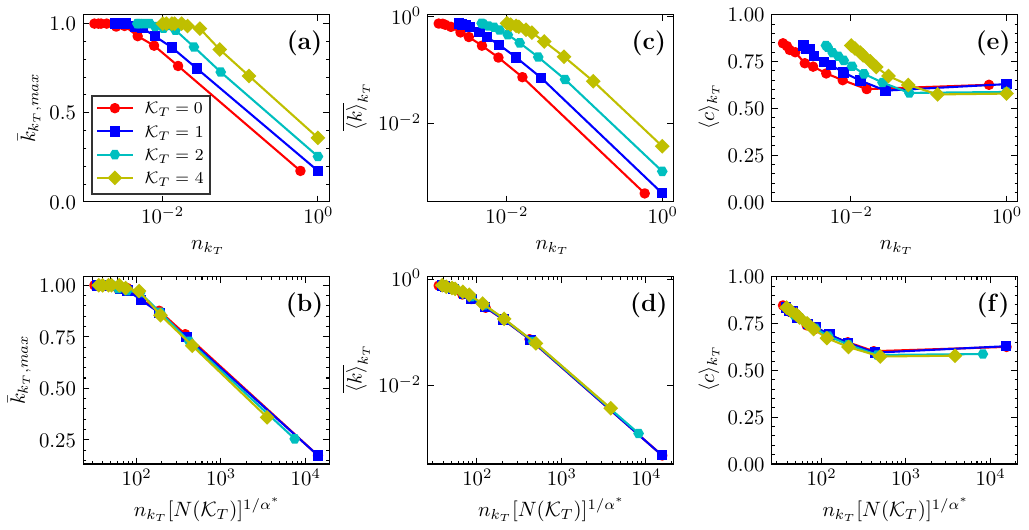}
\caption {{\color{blue}Scaling analysis of structural observables on the Internet ($\mathcal{K}_T = 0$) and its subgraphs $G(\mathcal{K}_T)$ (where $\mathcal{K}_T = 1, 2, 4$, the number of nodes corresponding to the three subgraphs is $N(\mathcal{K}_T) =$ 14135, 7510, and 3541, respectively). (a), (c), and (e) show the dependence of three structural observables on $n_{k_T}$, respectively. (b), (d), and (f) show the collapse plot after the horizontal axis is rescaled with the exponent $\alpha^* \approx 1$, respectively.}}
\label{Fig:8}
\end{center}
\end{figure*}

As shown in Fig.~S8, we observed that the scaling behavior of these dynamic observables is similar to that in Fig.~\ref{Fig:6}. Under the rescaling of Eq.~\eqref{Eq:3}, these curves of BA networks with different sizes could collapse onto the same master curve. In particular, an same scaling exponent, $\alpha = 1$, is produced here. Our results have important guiding significance for studying the dynamic behavior of large-scale networks. For instance, in practice, one possible outcome is to predict the synchronization stability of the initial large-scale network by taking the subnetwork as the study object and combining this scaling exponent. {\color{blue}Fig.~\ref{Fig:6} and Fig.~S8 show the results of BA networks under the parameter $m = 5$, and a consistent conclusion can be obtained when the parameter $m$ is set to other values (see Figs.~S9 and S10).}

{\color{blue}We perform the same study as Fig.~\ref{Fig:6} on CL networks with different heterogeneity exponent $\lambda$ (Figs.~S11 and S12 give the results for $\lambda=2.5$ and $3.5$, respectively)}, examining the dependence of scaling exponent $\alpha$ on the heterogeneity exponent $\lambda$, as shown in Fig.~\ref{Fig:7} (red solid circle, blue solid square, and green solid triangle). Our results show that, for each observable, the exponent $\alpha$ decreases gradually with the increase of the $\lambda$, which means that the exponent $\alpha$ strongly depends on the heterogeneity of the network. Moreover, for different observables, the exponent $\alpha$ has a large difference. Particularly, when $\lambda = 3$, each observable corresponds to an approximately same exponent, i.e., $\alpha \approx 1$, which is consistent with the results of BA scale-free networks.

When evaluating the scaling exponent $\alpha$ of synthetic networks, we can obtain this result by generating a series of networks with different initial sizes. For real networks, however, there is usually a single initial size. A natural question is how to calculate the scaling exponent for a particular network {\color{blue}with a single size}. To address this issue, we conduct further research on CL scale-free networks with a single initial size. {\color{blue}Specifically, we generate a CL scale-free network $G_0$ with $N_0=10^5$, and then obtain the subgraph $G(\mathcal{K}_T)$ under the given degree threshold $\mathcal{K}_T$. The $\mathcal{K}_T$ used here is essentially the same as the $k_T$ used above, with the main difference being that $\mathcal{K}_T$ is a degree threshold explicitly set for a single initial network, and its value range is usually limited to a very narrow interval. For example, as shown in Fig.~S13, the size of the initial network $G_0$ is $N_0=10^5$, its three subgraphs under degree thresholds $\mathcal{K}_T = 5, 10, 15$ are obtained, respectively, and the size of $G_(\mathcal{K}_T)$ is $N(\mathcal{K}_T)$ = 51734, 19901, and 10664, respectively. Next, we take $G_0$ and its three subgraphs as new initial reference networks to investigate their FSS behavior in DTR flows (see Fig.~S13 and Fig.~S14). Therefore, the subgraph size is within a reasonable range in this context. In other words, the subgraph size $N(\mathcal{K}_T)$ cannot be too close to $N_0$, nor too small. For simplicity, we chose the appropriate $\mathcal{K}_T$ so that the size of the subgraph is roughly equal to $N_0/2$, $N_0/5$, and $N_0/10$ (or $N_0/2$, $N_0/4$, and $N_0/8$). However, the $k_T$ can take on a wide range of values. For instance, for the BA scale-free network with $N_0=10^4$, when $k_T=100$, the size of the subgraph $G(k_T)$ is $N_{k_T}<100$, which is more than a hundred times smaller than the original network (see Fig.~\ref{Fig:3}).}

In the above context, $G_0$ and its subnetworks are taken as initial networks to study the FSS behavior of their observables in the DTR flow. By analogy with Eq.~\eqref{Eq:3}, under the FSS hypothesis, the observable $\mathcal{Z}$ satisfies,
\begin{equation}
\mathcal{Z}=f\left\{n_{k_{T}}\left[N\left(\mathcal{K}_{T}\right)\right]^{1/\alpha^*}\right\},
\label{Eq:4}
\end{equation}
where $N\left(\mathcal{K}_{T}\right)$ is the size of the subnetwork $G(\mathcal{K}_T)$, {\color{blue}and $n_{k_T}$ is the relative size of the remaining network after removing nodes $k \leqslant k_T$ from the subnetwork $G(\mathcal{K}_T)$. The behavior of the observable $\mathcal{Z}$ is completely determined by the exponent $\alpha^*$ and the scaling function $f$. Interestingly,} the scaling exponent of the initial network and its subnetwork is independent of the heterogeneity exponent $\lambda$ and shares the same scaling exponent with the BA network, namely, $\alpha^* \approx 1$, as shown in Fig.~\ref{Fig:7} (red hollow circle, blue hollow square, and green hollow triangle). {\color{blue}In particular, these three topological observables considered here correspond to an approximately identical exponent $\alpha^*$.} 

{\color{blue}Subsequently, we present a possible application: the exponent $\alpha^*$ is obtained by FSS analysis performed on the topological observables, then combined with this exponent to predict the dynamic characteristics of the initial network. The reason is that the pure calculation of the network's topology observables (such as the largest degree of the network) is usually a task with low computational complexity. In contrast, calculating dynamic observables (such as the eigenvalues of the network) is usually a task with high computational complexity. The significance of the scaling exponent, $\alpha^*$, is that we can predict the dynamic characteristics of the initial large-scale network with the help of the subgraph without spending extra time to calculate them separately. For example, based on $\alpha^*$ and the subnetwork $G(\mathcal{K}_T)$ of the initial large-scale network $G_0$, we can predict the behavior of the largest eigenvalue of the $G_0$ in the DTR flow, as shown in Fig.~S15.}

Finally, we confirm the universality of $\alpha^* = 1$ via real scale-free networks, and details of these real network datasets are shown in Table~\ref{table:1}. Fig.~\ref{Fig:8} shows the structural observables results of the Internet and its three subnetworks $G(\mathcal{K}_T)$, where $\mathcal{K}_T = 1, 2, 4$, and the results of dynamical observables are shown in Fig.~S16. {\color{blue}To ensure the subnetwork $G(\mathcal{K}_T)$ has approximately the same topology properties as the original network (see Figs.~S17-S19), the value of $\mathcal{K}_T$ should not be selected too large.} Then we take the Internet and its three subnetworks as the research object, and perform DTR on them, respectively. By rescaling the horizontal axis according to Eq.~\eqref{Eq:4}, these observable curves of subnetworks and the original Internet network can collapse onto the same master curve, where the exponent $\alpha^* \approx 1$. For other real scale-free networks, approximately the same exponent $\alpha^*$ is obtained, as shown in Table~\ref{table:1}. The result confirms that scale-free networks and their subnetworks share the same universality class exponent.

\subsection{Discussion}\label{3.3}
{\color{blue}Fig.~\ref{Fig:6} and Fig.~\ref{Fig:7} show that the corresponding scale exponent $\alpha$ of the BA and CL synthetic network is inconsistent. Next, we give the possible reasons: the CL network is a static model, and its connection mode differs from the preferential attachment growth mode of the BA network. For the BA model, the degree increases with the age of nodes, while the DTR tends to delete the most recent nodes. Therefore, DTR can approximately return the BA network to an earlier state, implying that the artificially generated smaller-size BA network can be approximately used as a nested subgraph of the larger-size BA network. Indeed, the degree distribution shows that the scale-free network of BA with different sizes is self-similar, as shown in Fig.~S20(a). This result means that the scaling exponent obtained in the DTR flow is approximately the same ($\alpha \approx \alpha^* \approx 1$) whether BA networks with different sizes or the DTR snapshots are used as the original network. However, the degree distribution of CL models with different sizes has certain differences, especially in the tail [as shown in Fig.~S20(b)]. This indicates that the artificially generated smaller-size CL network cannot be used as the nested subgraph of the larger-size CL network. In contrast, the subgraph extracted by DTR can be approximately used as the nested subgraph of the initial large-scale CL network [see Fig.~S1(a)]. Therefore, exponents $\alpha$ and $\alpha^*$ are different for CL models. In conclusion, the results of the hollow symbol reported in Fig.~\ref{Fig:7} and the results of the real network produced in the same way in Fig.~\ref{Fig:8} describe a feature of finite-size self-similar networks in DTR flows from another perspective.
 
Furthermore, the BCR technique proposed by song {\it et al.}~\cite{Song2005} connects the representation of dimensionality with the definition of distance. They use the shortest path length between nodes to define the similar distance for the network's renormalization and define the fractal dimension of the network. Recently, considering the ultrasmall-world property of real-world networks, which cannot provide a wide range of distance values. Within the context of hidden metric spaces~\cite{Serrano2008}, the GR~\cite{Garcia2018} technique comes into being, this method has achieved excellent results in most scenarios, which is further confirmed in a series of literature and publications~\cite{zheng2021,boguna2021,serrano2021,almagro2022}. The scaling function employed in this paper is similar in spirit to the Refs.~\cite{Radicchi2008,Radicchi2009}, by comparison, our work is performed on the DTR technique. In short, all these renormalization schemes contain the idea of dimension reduction, reducing dimension redundancy, which is of great engineering significance for processing massive high-dimensional datasets.

In this work, we employ scale-free networks as research objects, and these network degree values are usually distributed in a relatively wide range. However, their degree values are usually distributed in a relatively narrow range for other homogeneous networks, such as ER random networks, the national highway network with homogeneous topology, or the European power grid with exponential degree distribution. For these networks, a small degree threshold may induce a subgraph with few nodes, which may bring inconvenience related research. On the other hand, considering that the node's degree is a local feature of the network, the information provided by it is often limited or incomplete to a certain extent. Therefore, it may be worth studying other network features as an alternative to the degree threshold. Under these circumstances, developing a subgraph extraction or coarse-graining procedure suitable for more scenarios is the direction we still need to work hard in the future.}

\section{Conclusion}\label{4}
{\color{blue}In this paper, we study the statistical properties of some representative observables of scale-free networks in the DTR flow, and experimental studies show that these observables obey the scaling law. First, we find that some properties of networks show scaling behavior in DTR flows.} 
Furthermore, our results also show that some observables deviate from the pure scaling behavior in DTR flows due to the finite-size effect of networks. Interestingly, in the context of FSS analysis, we find a scaling function that always yields observed data collapse at different network sizes. {\color{blue}Specifically, for BA scale-free networks with different initial sizes, the FSS confirmed this universal scaling law and revealed a scaling function with a single exponent $\alpha \approx 1$ to capture the behavior of observables in DTR flows. For CL scale-free networks with different initial sizes, our results show that the scaling exponent depends on its degree distribution exponent $\lambda$. Notably, for CL or real-world scale-free network with a single initial size, our results show that the observable of the initial network and its subnetworks share the same scaling exponent as the BA network.} These results suggest that subnetworks could also be applied to perform finite-size scaling and the scaling exponent can be determined from a single snapshot of the topology. From the perspective of the application, the scaling exponent obtained here can be used as the foundation for predicting the structure and dynamic characteristics of large-scale networks, which has important guiding significance for reducing the time complexity of the large-scale numerical simulation.

\bibliographystyle{IEEEtran}
\bibliography{FSS_DTR}

\end{document}